\documentclass[11pt,a4paper]{article}
\usepackage{geometry}
\usepackage[english]{babel}
\usepackage{amsmath,amsthm}
\usepackage{amsfonts}
\usepackage{graphicx}
\usepackage{bm}
\usepackage[colorlinks,linkcolor=blue,anchorcolor=blue,citecolor=blue,urlcolor=blue,filecolor=blue,menucolor=blue,runcolor=blue]{hyperref}
\usepackage{ulem}
\usepackage{authblk}

\begin{document}

\title{Evidence for nematic superconductivity of topological surface states in PbTaSe$_2$}

\author[1]{Tian Le}
\author[2]{Yue Sun \thanks{Corresponding author: sunyue@phys.aoyama.ac.jp}}
\author[3]{Hui-Ke Jin}
\author[1]{Liqiang Che}
\author[1]{Lichang Yin}
\author[1]{Jie Li}
\author[1]{G. M. Pang}
\author[4]{C. Q. Xu}
\author[5]{L. X. Zhao}
\author[6]{S. Kittaka}   
\author[6]{T. Sakakibara}   
\author[7]{K. Machida}   
\author[8]{R. Sankar}
\author[1,9,10]{H. Q. Yuan}
\author[5,11]{G. F. Chen}
\author[4]{Xiaofeng Xu \thanks{Corresponding author: xiaofeng.Xu@cslg.edu.cn}}
\author[12,10]{Shiyan Li}
\author[3,5,10,13]{Yi Zhou}
\author[1,9,10]{Xin Lu \thanks{Corresponding author: xinluphy@zju.edu.cn}}

\affil[1] {\textit{Center for Correlated Matter and Department of Physics, Zhejiang University, Hangzhou 310058, China}}
\affil[2] {\textit{Department of Physics and Mathematics, Aoyama Gakuin University, Sagamihara 252-5258, Japan}}
\affil[3] {\textit{Department of Physics, Zhejiang University, Hangzhou 310027, China}}
\affil[4] {\textit{Advanced Functional Materials Lab and Department of Physics, Changshu Institute of Technology, Changshu 215500, China}}
\affil[5] {\textit{Beijing National Laboratory for Condensed Matter Physics, Institute of Physics, Chinese Academy of Sciences, Beijing 100190, China}}
\affil[6] {\textit{Institute for Solid State Physics (ISSP), The University of Tokyo, Kashiwa, Chiba 277-8581, Japan}}
\affil[7] {\textit{Department of Physics, Ritsumeikan University, Kusatsu, Shiga 525-8577, Japan}}
\affil[8] {\textit{Institute of Physics, Academia Sinica, Nankang, Taipei 11529, Taiwan}}
\affil[9]  {\textit{Zhejiang Province Key Laboratory of Quantum Technology and Device, Zhejiang University, Hangzhou 310027, China}}
\affil[10] {\textit{Collaborative Innovation Center of Advanced Microstructures, Nanjing University, Nanjing, 210093, China}}
\affil[11] {\textit{Collaborative Innovation Center of Quantum Matter, Beijing 100084, China}}
\affil[12] {\textit{State Key Laboratory of Surface Physics, Department of Physics, and Laboratory of Advanced Materials, Fudan University, Shanghai 200433, China}}
\affil[13]{\textit{CAS Center for Excellence in Topological Quantum Computation, University of Chinese Academy of Sciences, Beijing 100190, China}}

\date{}

\maketitle
\textbf{Spontaneous symmetry breaking has been a paradigm to describe the phase transitions in condensed matter physics. In addition to the continuous electromagnetic gauge symmetry, an unconventional superconductor can break discrete symmetries simultaneously, such as time reversal and lattice rotational symmetry. In this work we report a characteristic in-plane 2-fold behaviour of the resistive upper critical field and point-contact spectra on the superconducting semimetal PbTaSe$_2$ with topological nodal-rings, despite its hexagonal lattice symmetry (or $D_{3h}$ in bulk while $C_{3v}$ on surface, to be precise). However, we do not observe any lattice rotational symmetry breaking signal from field-angle-dependent specific heat. It is worth noting that such surface-only electronic nematicity is in sharp contrast to the observation in the topological superconductor candidate, Cu$_x$Bi$_2$Se$_3$, where the nematicity occurs in various bulk measurements. In combination with theory, superconducting nematicity is likely to emerge from the topological surface states of PbTaSe$_2$, rather than the proximity effect. The issue of time reversal symmetry breaking is also addressed. Thus, our results on PbTaSe$_2$ shed new light on possible routes to realize nematic superconductivity with nontrivial topology.}

In analogy to nematic liquid crystals, electronic nematicity \cite{Fradkin} breaks the lattice rotational symmetry into 2-fold and has been widely observed in various condensed matter systems, such as cuprates, iron-based superconductors, and heavy fermions \cite{Kivelson:98, sato2017thermodynamic, okazaki2011rotational, Kasahara:12, ronning2017electronic, fernandes2014drives}, whose origins still remain controversial for each case. On the other hand, together with topological insulators and semimetals, topological superconductors (TSCs) have been attracting more and more attentions \cite{RevModPhys.83.1057, RevModPhys.82.3045, 031214-014501, 0034-4885-80-7-076501}, partially due to their promising applications on topological quantum computation \cite{Kitaev:03,Nayak:08}. Among them, lattice rotational symmetry has been discovered to be broken in some doped topological superconductors such as Cu$_x$Bi$_2$Se$_3$ \cite{Hor:10}, where a 2-fold oscillation is reported in its superconducting state by nuclear magnetic resonance (NMR) and specific heat measurements, complying with an anisotropic but fully-gapped superconducting order parameter in an odd-parity $E\rm_u$ symmetry \cite{PhysRevB.90.100509,PhysRevB.94.180504, Yonezawa2017, Matano2016, PhysRevX.7.011009, Cheneaat1084, PhysRevX.8.041024}. In this case, the nematicity is believed to be associated with the bulk topological superconductivity (SC) in Cu$_x$Bi$_2$Se$_3$. Nematic superconductivity has been observed in Cu-, Nb-, and Sr-doped Bi$_2$Se$_3$ compounds, which exhibit 2-fold symmetry in field-rotational specific heat, magnetic torque, upper critical field, magnetization, and STM spectra in their superconducting states~\cite{Yonezawa2017,PhysRevX.7.011009,Pan:16,Du:17,Andersen:18,Shen:17,Tao:18}. Search for nematic superconductors in other types of TSC candidates is challenging but desirable in order to explore the intricate relationship between nematicity and nontrivial topology of quasi-particle bands \cite{PhysRevB.90.100509,PhysRevB.94.180504,PhysRevLett.105.097001}.

Recently, the existence of topological nodal-rings, an indirect gap opened by the spin-orbit coupling and associated topological surface states in the hexagonal $D_{3h}$ lattice compound PbTaSe$_2$ has been established by a comparative study on density functional theory (DFT) calculations and angle-resolved photoemission spectroscopy (ARPES) measurements \cite{bian2016topological,PhysRevB.93.245130}. PbTaSe$_2$ can be viewed as an alternating stacking of Pb layers and hexagonal TaSe$_2$ units while the crystal lattice still remains ditrigonal-dipyramidal symmetry as shown in Fig. 1(a), even though the Ta atom is off the crystal inversion center. Bulk superconductivity in PbTaSe$_2$ has been reported with $T$$\rm_c$ around 3.7 K and a fully gapped spin-singlet superconducting state has been claimed by NMR, thermal conductivity and penetration depth studies on PbTaSe$_2$ \cite{PhysRevB.97.184510, PhysRevB.93.020503, PhysRevB.93.060506}. The scanning tunneling microscopy (STM) observations suggest that the topological surface states in PbTaSe$_2$ display a nodeless SC with probable Majorana zero-energy modes in its vortex cores \cite{PhysRevB.89.020505, Guane1600894}. As a chemically stoichiometric compound, PbTaSe$_2$ thus serves as a promising candidate of pure TSCs.

In this article, based on electrical resistivity and point-contact spectroscopy (PCS) results on single crystalline PbTaSe$_2$ with $T$$\rm_c =$ 3.8 K, we report an unambiguous 2-fold symmetry of the in-plane upper critical field and the PCS conductance curves, when the magnetic field is rotated in the $ab$ plane. Our results imply that its in-plane superconducting gap minima are always located along Se-Se (Pb-Pb) bonds below $T\rm _{c}$ as well as the resistive upper critical field $H\rm^{R}_{c2}$. However, the nematic behaviour is absent in the bulk measurements of field-rotational specific heat without any 2-fold trace. We argue that the observed nematicity is most likely attributed to the superconducting topological surface states rather than the bulk SC in PbTaSe$_2$.

Electrical resistivity measurements on PbTaSe$_2$ have been conducted in a Corbino-shape-like configuration as illustrated in Fig. 1(b) to  keep the current direction always parallel to $c$-axis and perpendicular to the magnetic filed during its rotation in the $ab$-plane. The insets of Fig. 1(c) show a resistive superconducting transition at $T\rm_{c}$ $\sim$ 3.8 K under zero-field and an upper critical field $H\rm^{R}_{c2}$ $\sim$ 1.0 T at 1.8 K for the field in the $ab$ plane. The field dependence of electrical resistivity at different angles in the $ab$-plane are summarized in Fig. 1(d) as a contour plot, where one of the Se-Se (Pb-Pb) bond directions is defined as the $x$-axis from the Laue diffraction pattern and $\phi$ is the azimuth angle between the field direction and $x$-axis. The zero resistance in the superconducting state becomes normal at distinct field magnitudes in different field-angles, and there exists a clear 2-fold oscillation for the upper critical field as indicated by the dashed line in Fig. 1(d) (Refer to the original data in the Supplementary Figure S1). The dashed line corresponds to a fitting function $H\rm^{R}_{c2} (\phi)$ = $H{\rm_0}$+$H{\rm_2}{\rm{cos}}(2\phi)$+$H{\rm_6}\rm{cos}(6\phi)$ with $H\rm_0$ = 0.9 $\pm$ 0.02 T, $H\rm_2$ = -0.18 $\pm$ 0.02 T, $H\rm_6$ = 0.01 $\pm$ 0.005 T. The 2-fold oscillation amplitude $H\rm_2$ is as large as 20$\%$ of $H\rm_0$ whereas the 6-fold component $H\rm_6$ is only 1$\%$ of $H\rm_0$, implying a strong 2-fold anisotropy of the upper critical field as a single nematic domain in the PbTaSe$_2$ crystals. This 2-fold behaviour can not be ascribed to vortex motions because the current is always along $c$-axis for this Corbino-shape configuration. Interestingly, the field direction at the minimum $H\rm^{R}_{c2}$ is close to one of the Se-Se (Pb-Pb) bonds, which is probably determined by its intrinsic superconducting properties such as the anisotropic SC gap. Considering the $D_{3h}$ symmetry of PbTaSe$_2$ lattice, the two-fold symmetry of in-plane $H\rm_{c2}$ strongly suggests a rotational symmetry breaking in the superconducting state and supports an electronic nematicity in PbTaSe$_2$ \cite{PhysRevB.90.100509,PhysRevB.94.180504,PhysRevLett.105.097001}.

\begin{figure}
\includegraphics[angle=0,width=0.99\textwidth]{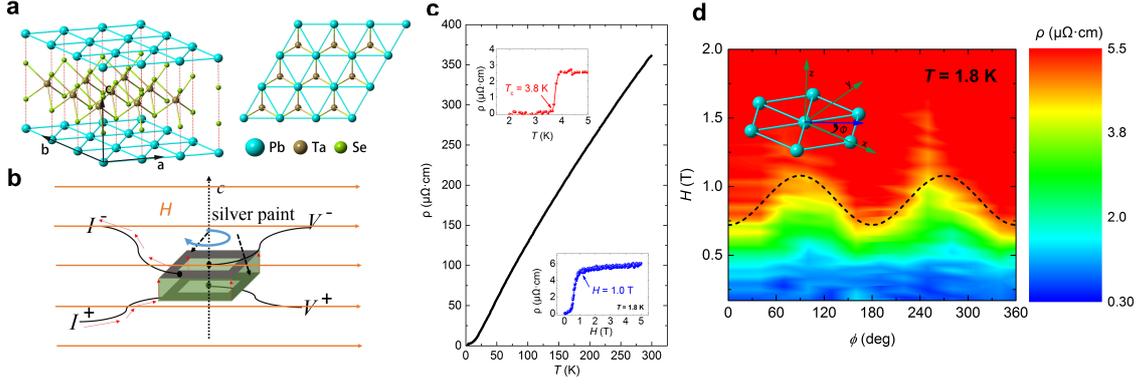}
\vspace{-12pt} \caption{\label{Figure1}\textbf{Evidence of 2-fold symmetric behaviour in PbTaSe$_2$ from electrical resistivity.} \textbf{(a)} The side view (left) and top view (right) of PbTaSe$_2$ crystal structure. \textbf{(b)} Schematic illustration of the Corbino-shape like electrode configuration on PbTaSe$_2$ with silver paint. \textbf{(c)} Temperature-dependent electrical resistivity from 300 K to 1.8 K for PbTaSe$_2$ with the electrode configuration in (b). Insets show the resistive transition at the superconducting $T\rm_c \sim$ 3.8 K and $H\rm^{R}_{c2} \sim $ 1.0 T at 1.8 K. \textbf{(d)} A color contour plot of the electrical resistivity for different magnetic fields at different azimuthal angles $\phi$ in the $ab$-plane with the Corbino-shape like electrode configuration, where $\phi$ is defined as the angle deviating from Se-Se (Pb-Pb) bond as shown in the inset.}
\vspace{-12pt}
\end{figure}

Field-rotational soft point-contact spectroscopy has been applied to investigate the possible nematic superconductivity in PbTaSe$_2$ \cite{PhysRevB.83.104519, 0953-2048-23-4-043001}, where the junction current across the normal metal-superconductor interface is along $c$-axis for each contact on PbTaSe$_2$ as shown in Fig. 2(a). The temperature and magnetic field dependence of the zero-bias resistance (ZBR) for the point-contact on PbTaSe$_2$ is shown in Fig. 2(b), confirming a similar superconducting transition temperature $T\rm_c$ in zero-field and upper critical field $H\rm^{R}_c$ at 1.8 K as in the electrical resistivity measurements on PbTaSe$_2$. Figure 2(c) displays a representative normalized PCS conductance curve as a function of biased voltage at 1.8 K with a double-peak feature at $\sim \pm$0.6 meV. The conductance curves do not show any dip structure at high bias voltages, ensuring the ballistic nature of our PCS contacts on PbTaSe$_2$ \cite {doi:10.1063/1.5030447, PhysRevB.69.134507}.

\begin{figure}
\includegraphics[angle=0,width=0.99\textwidth]{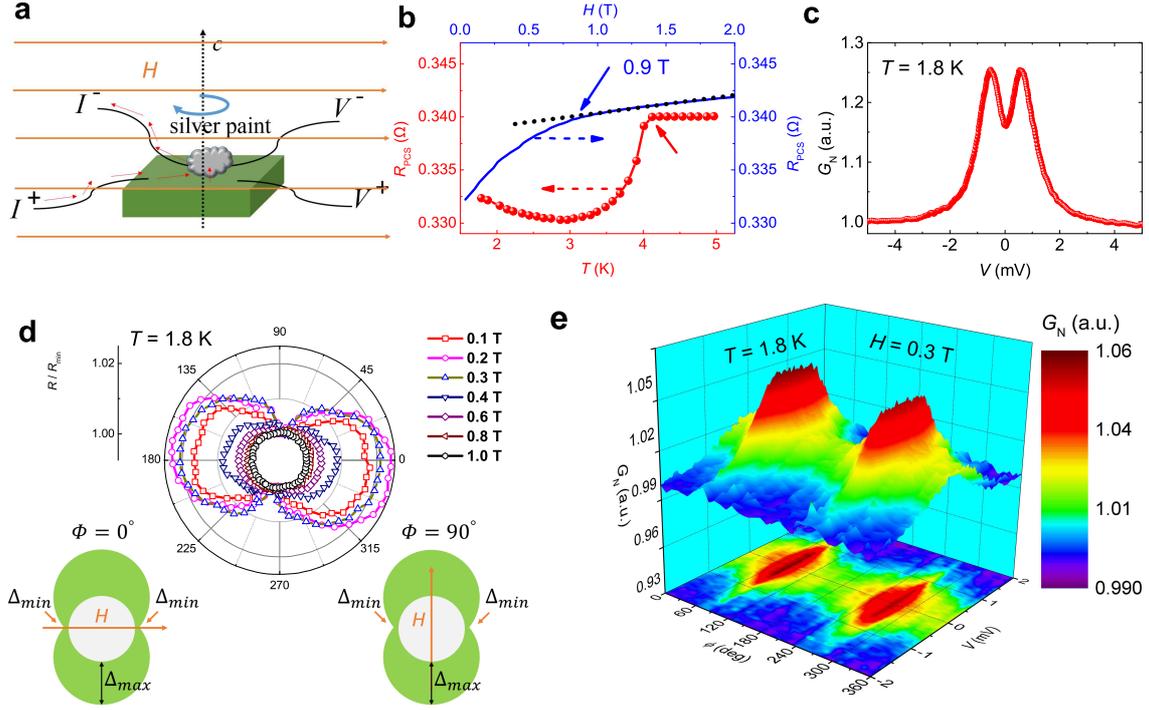}
\vspace{-12pt} \caption{\label{Figure2} \textbf{Evidence of 2-fold symmetric behaviour in PbTaSe$_2$ from soft point-contact spectroscopy} \textbf{(a)} Schematic illustration of a soft point-contact on PbTaSe$_2$. \textbf{(b)} Temperature and field dependence of zero-bias PCS resistance, showing a consistent $T\rm_c \sim$ 3.8 K and $H\rm^{R}_{c2} \sim $ 0.9 T at 1.8 K. \textbf{(c)} A representative soft-PCS conductance curve at 1.8 K, showing a typical double-peak structure. \textbf{(d)} Azimuthal plot of the angle dependence of zero-bias resistance for the soft PCS in different magnetic fields within the $ab$-plane at 1.8 K. \textbf{(e)} Three-dimensional plot of soft PCS conductance curves as a function of bias voltage and field angle at 1.8 K and 0.3 T.}
\vspace{-12pt}
\end{figure}

The azimuthal angle dependence of zero-bias resistance of point-contact on PbTaSe$_2$ at 1.8 K is plotted in Fig. 2(d), where the magnetic field is rotated in the $ab$-plane with a series of field sizes. A clear 2-fold symmetry of the ZBR is present in Fig. 2(d) with a dumbbell-shape below $H\rm^{R}_{c2}$$\sim$ 1.0 T, probably signaling an anisotropic superconducting gap with a $C_2$ symmetry (More examples of the 2-fold symmetry behaviour for PCS on PbTaSe$_2$ are included in Supplementary Figure S2). The magnitude of the 2-fold oscillation smoothly evolves with field and reaches as high as 2\%, while the direction of the maximum ZBR is always close to one of the Se-Se (Pb-Pb) bonds with deviations less than 10 degrees. This 2-fold symmetry is observed only in the superconducting state and the field-rotational ZBR becomes isotropic in the normal state either above $H\rm^{R}_{c2}$ at 1.8 K or above $T_c$ in zero-field. A misalignment of the PbTaSe$_2$ crystal's $ab$-plane with the field direction should cause a two-fold oscillation in the normal state (see supplementary Fig. S4) because there exists a large anisotropic magnetoresistance between the $c$ and $ab$ plane. Since the 2-fold symmetry is absent in the normal state, we thus argue for its intrinsic origin observed only in the superconducting state for our soft point-contacts on PbTaSe$_2$. A three-dimensional contour plot of the PCS conductance curves $G(V)$ at 1.8 K and 0.3 T as a function of field directions is shown in Fig. 2(e), which also displays a well-defined 2-fold symmetry for both the enhanced conductance peak intensity and width with the presence of Andreev reflection (The double-peak structure in zero-field has been smeared into a single zero-bias peak by the magnetic field with $H=0.3$ T). When the field is aligned in the $x$-axis parallel to Se-Se (Pb-Pb) bonds, the superconducting gap is greatly reduced by field and the conductance curve $G$($V$) has a much weaker and thinner feature, consistent with the $H\rm^{R}_{c2}$ anisotropy observed in resistivity measurements. In the case of our PCS measurements, the junction current is also along $c$ axis for contacts on the $ab$ plane of PbTaSe$_2$ and the influence of vortex motion can be excluded. Therefore, the observed 2-fold symmetry behaviour in the superconducting state strongly favors a nematic superconductivity of PbTaSe$_2$ with lattice rotational symmetry breaking.

In sharp contrast, no trace of any symmetry oscillation is present in the field-rotational specific heat measurements on PbTaSe$_2$ for its superconducting state as shown in Fig. 3(a). We thus conclude the absence of nematicity in the bulk superconducting PbTaSe$_2$ with an isotropic gap. In order to address this discrepancy between the bulk specific heat and resistivity or PCS results, it is interesting to note a systematic difference of the upper critical field in PbTaSe$_2$ between the resistivity and specific heat results, which give $H\rm^{R}_{c2}$ $\sim$ 1.0 T as in Fig. 2(b) and $H\rm^{HC}_{c2}$ $\sim$ 0.2-0.3 T at 0.35 K for field in the ab plane as in Fig. 3(b) (The polar-angle $\theta$ independence of $\Delta C$/T between 0.2-0.3 T indicates the total suppression of SC in the $ab$-plane). Since specific heat measurements can only probe the bulk SC whereas electrical resistivity or PCS can detect both bulk and surface SC, such a discrepancy indicates that the bulk SC has already been destroyed by a magnetic field of 0.2-0.3 T while the surface SC can survive even up to 1.0 T, yielding a distinct in-plane upper critical field (Refer to more discussions in the supplementary note 3 to exclude other scenarios for different $H\rm_{c2}$). We notice that the field dependence of the PCS excessive current at 0.3 or 1.8 K, defined as the integration of PCS conductance $G(V)$ subtracted by the normal state baseline, shows a kink structure around $H\rm^{HC}_{c2} \sim$ 0.2 T and persists up to $H\rm^{R}_{c2}$. Its field evolution deviates from the common behaviour shared by other conventional superconductors as shown in Fig. 3(d) \cite{PhysRevB.73.134508,Huang_2018} (The original field dependence of PCS curves are included in the Supplementary Figure S5), suggesting the contribution of superconducting surface states to the soft PCS spectra in addition to the bulk. More detailed arguments on the presence of superconducting topological surface states can be found in the Supplementary Figure S6 \& S7. Our angle-dependent ZBR of point-contact spectra displays the largest nematic component at 0.2 T and it persists up to 0.8 T, much higher than $H\rm^{HC}_{c2}$, favoring that the nematic behaviour originates from the surface superconductivity rather than bulk SC of PbTaSe$_2$. On the other hand, if there were multiple fine domains with random nemtaic orientations in the bulk, the nematicity would be averaged out in the field-rotational specific heat measurements, resulting in no symmetry oscillation in the field-angle dependence of the specific heat. However, such randomly-oriented nematic domains would also average out the resistivity, and can not reconcile with the observation of a clear 2-fold upper critical field $H_{c2}^{R}$ in the resistive measurements. Therefore we conclude that nematic superconductivity is absent in the bulk while it emerges on the surface. To the best of our knowledge, it is the first time that surface-only nematic superconductivity is observed experimentally.

\begin{figure}
\includegraphics[angle=0,width=0.99\textwidth]{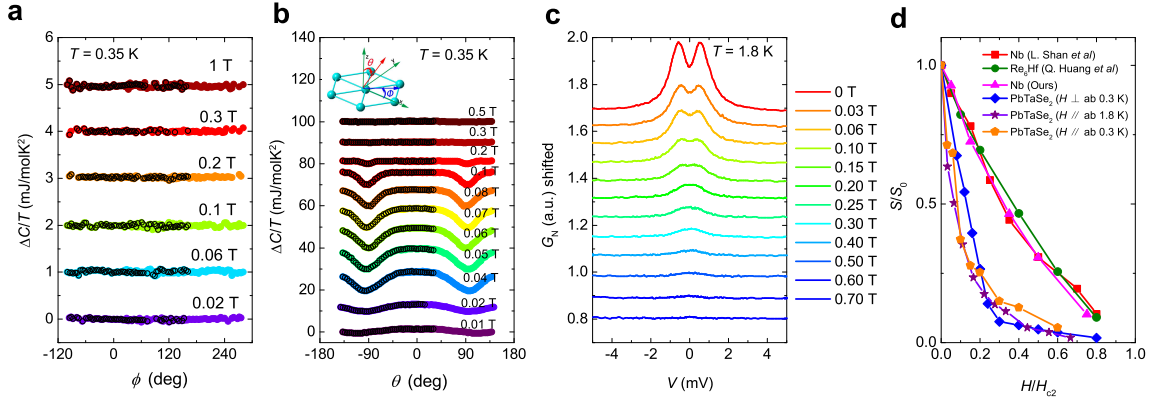}
\vspace{-12pt} \caption{\label{Figure3}\textbf{No oscillation of bulk superconductivity in PbTaSe$_2$ and the signature of topological surface superconductivity} \textbf{(a) and (b)} Azimuthal- and polar- angle dependence of specific heat of PbTaSe$_2$ for the field rotated in the $ab$ and $ac$ planes at 0.35 K, respectively. $\Delta C(\phi)/T$ is defined as $[C(\phi)-C(0^\circ)]/T$, and each subsequent curve is shifted vertically by 1 mJ/mol-K$^2$. $\Delta C(\theta)/T$ is defined as $[C(\theta)-C(-90^\circ)]/T$, and each subsequent curve is shifted vertically by 10 mJ/mol-K$^2$. Points in black are the measured data and other points are the mirror data. \textbf{(c) and (d)} Field evolution of the soft PCS conductance curves and excessive currents at 0.35 K.} 
\vspace{-12pt}
\end{figure}

Regarding symmetries, such surface-only nematic superconductivity should originate from the topological surface states themselves instead of the proximity effect. DFT calculations and ARPES measurements revealed that the topological surface states form two helical Fermi surfaces (FSs) around the $\bar{\Gamma}$ point~\cite{bian2016topological,PhysRevB.93.245130}. 
With the help of the Ginzburg-Landau theory and a microscopic two-channel model, we are able to study theoretically what possible SC states may emerge from the helical surface states in PbTaSe$_2$ \cite{jin2019theory}, and found: (1) nematic SC states are favored by the inter-site attraction; (2) for a time-reversal invariant SC state with nematicity, its SC gap must be nodal; (3) in the absence of spin-singlet and -triplet admixture in the SC pairing, the quasiparticle gap must be of hexagonal symmetry in a fully gapped SC state. Considering the full SC gap observed by STM \cite{Guane1600894}, we are able to deduce that the time-reversal symmetry (TRS) must be broken in such a surface SC state with strong spin-singlet and -triplet mixing, and the resulting state is topologically nontrivial and will be able to host Majorana zero modes in vortex cores. The winding number of the SC, $N_{s} = \frac{1}{2\pi}\oint_{\text{FS}} d\bm{k}\cdot \nabla_{\bm{k}}\arg \Delta_{s}(\bm{k})$, takes different values on two surface FSs, namely, either $N_{+}=1,N_{-}=3$ or $N_{+}=-3,N_{-}=-1$, where $s=\pm$ refers to the outer or inner FS and $\Delta_s({\bm{k}})$ is the corresponding SC pairing function. Notably, such a TRS breaking state is qualitatively different from the TSC induced by the proximity effect on a 3D strong topological insulator surface as proposed by Fu and Kane \cite{PhysRevLett.100.096407} and realized in the Bi$_2$Te$_3$/NbSe$_2$ hetero-structure \cite{Wang52}. The latter does not break TRS indeed.

Finally, we would like to emphasize that our proposed surface nematic superconductivity in PbTaSe$_2$ is distinctive from the odd-parity TSC candidate Cu$_x$Bi$_2$Se$_3$: Even though a similar rotational symmetry breaking was reported in Cu$_x$Bi$_2$Se$_3$, its nematicity is rather associated with the bulk SC in an $E_u$ gap symmetry \cite{PhysRevB.90.100509}. When the bulk SC in PbTaSe$_2$ is totally suppressed by magnetic field, its spin-polarized topological surface SC still survives, which can't be simply ascribed to SC proximity effect as in the Bi$_2$Se$_3$/NbSe$_2$ heterostructure \cite{Wang52}. The exact origin of the surface nematic SC in PbTaSe$_2$ still remains unresolved, however, our results suggest that this chemically stoichiometric superconductor PbTaSe$_2$ should serve as a unique platform to study the intricate relationship between the nematic and topological SC. More experimental measurements, including low-temperature ARPES and STM, are desirable to identify the exact nature of the nematic surface SC in PbTaSe$_2$. It is also interesting to study whether the surface SC in PbTaSe$_2$ has broken the time reversal symmetry by polar Kerr measurements, whereas it is evidenced to be conserved in the bulk SC by the $\mu$SR experiment on PbTaSe$_2$.

\textbf{Methods}

\textbf{Sample preparation.}
Single crystals of PbTaSe$_2$ were grown by the chemical vapor transport methods as described in Ref. \cite{0953-8984-29-9-095601} and Ref. \cite{0256-307X-33-3-037401}. The crystal orientations were determined by X-ray Laue diffraction. The sample is cleaved at ambient conditions and electrodes are made within ten minutes for resistivity and point-contact spectroscopy measurements.

\textbf{Resistivity measurement.}
Electrical resistivity of PbTaSe$_2$ was measured by the conventional four-probe method with a Corbino shape to guarantee the current always perpendicular to the $ab$ plane. Electrodes are made with silver paint (SPI05001-AB), which can be dry within several minutes.

\textbf{Point-contact spectroscopy measurement.}
Soft point-contacts on PbTaSe$_2$ were prepared by attaching a 30 $\mu$m diam gold wire with a silver-paint drop at the end on the freshly-cleaved surface at room temperature. In such a configuration, thousands of parallel nanoscale channels are assumed between individual silver particles and the crystal surface. Mechanical point-contacts on PbTaSe$_2$ in a needle-anvil style were prepared by engaging an electrochemically-etched sharp gold tip on sample surface by piezo-controlled nano-positioners. The conductance curves as a function of bias voltage, $G$($V$), were recorded with the conventional lock-in technique in a quasi-four-probe configuration. The output current is mixed with dc and ac components, which were supplied by the model 6221 Keithley current source and model 7265 DSP lock-in amplifier, respectively. The first harmonic response of the lock-in amplifier is proportional to its point-contact resistance d$V$/d$I$ as a function of the biasd voltage $V$.

\textbf{Heat capacity measurement.}
The field-orientation dependence of the specific heat was measured in an 8 T split-pair superconducting magnet with a $^3$He refrigerator to cool the samples to temperatures below 300 mK. Without thick pumping tube outside, the refrigerator insert could be easily rotated using a stepper motor mounted at the top of a magnet Dewar. The overall angular resolution of the field direction is better than 0.01 deg. The hexagonal $ab$ plane of the sample can be set either parallel or perpendicular (with a rectangular accessory) to the horizontal magnetic field to measure the azimuthal angle ($\phi$) or the polar angle ($\theta$) dependence of the specific heat \cite{PhysRevB.96.220505}.

\textbf{Low temperature measurements.}
Angle-dependent resistivity and soft point-contact spectroscopy down to 1.8 K were measured in a Quantum Design Physical Property Measurement System (9T-PPMS) equipped with a sample rotator. Mechanical and soft point-contact spectroscopy down to 0.3 K were measured in Oxford $^3$He cryostat with an 8 T magnet.

\textbf{Theoretical calculation.}
Superconducting pairing symmetries were analyzed by group theory, Green's function, as well as the linearized gap equation method.

\textbf{Acknowledgements}
We are grateful for valuable discussions with C. Ren, H. Zhen and Y. Liu. Work at Zhejiang University is supported by National Key R \& D Program of China (2016FYA0300402 and 2017YFA0303101) and the National Natural Science Foundation of China (NSFC) (Grant No. 11674279, 11374257). X.F.X. is supported in part by the NSFC Grant No. U1732162. X.L. would like to acknowledge support from the Zhejiang Provincial Natural Science Foundation of China (LR18A04001). H.K.J. and Y.Z. are supported in part by National Key Research and Development Program of China (No.2016YFA0300202), National Natural Science Foundation of China (No.11774306), and the Strategic Priority Research Program of Chinese Academy of Sciences (No. XDB28000000). The present work was partly supported by KAKENHI (19K14661, 15H05883, 18H01161, and JP17K05553) from JSPS, and ``J-Physics'' (18H04306).




\end{document}